\documentclass[prl,reprint,twocolumn,amsmath,amssymb,floatfix,showpacs]{revtex4}
\usepackage{hyperref}
\usepackage{graphicx} 
\usepackage{amsmath}
\usepackage[dvips]{epsfig}
\newcommand{\beq}{\begin{equation}}
\newcommand{\eeq}{\end{equation}} 
\newcommand{\beqa}{\begin{eqnarray}}
\newcommand{\eeqa}{\end{eqnarray}}
\newcommand{\ba}{\begin{array}}
\newcommand{\ea}{\end{array}}

\begin{document}

\title{First observation of bright solitons in bulk superfluid $^4$He}

\author{Francesco Ancilotto$^1$, David Levy$^2$, Jessica Pimentel$^3$, and Jussi Eloranta$^3$}

\affiliation{$^1$Dipartimento di Fisica e Astronomia ``Galileo Galilei''
and CNISM, Universit\`a di Padova, via Marzolo 8, 35122 Padova, Italy and CNR-IOM Democritos, via Bonomea, 265 - 34136 Trieste, Italy\\
$^2$Department of Physics and Astronomy, California State University at Northridge, California 91330, USA\\
$^3$Department of Chemistry and Biochemistry, California State University at Northridge, California 91330, USA}

\begin{abstract}
The existence of bright solitons in bulk superfluid $^4$He is demonstrated by time-resolved shadowgraph imaging experiments and density functional theory (DFT) calculations. The initial liquid compression that leads to the creation of non-linear waves is produced by rapidly expanding plasma from laser ablation. After the leading dissipative period, these waves transform into bright solitons, which exhibit three characteristic features: dispersionless propagation, negligible interaction in two-wave collision, and direct dependence between soliton amplitude and the propagation velocity. The experimental observations are supported by DFT calculations, which show rapid evolution of the initially compressed liquid into bright solitons. At high amplitudes, solitons become unstable and break down into dispersive shock waves.
\end{abstract} 
\date{\today}

\pacs{67.25.D-,67.25.bf, 67.85.dt}

\maketitle

Solitons are localized non-linear waves in a medium, which do not disperse as a function of time and exhibit no interaction during a two-wave collision. After their discovery in the early eighteen hundreds, solitons have been observed in many different media, which exhibit pronounced non-linear response. In recent years, solitons have become an intense field of research due to their important applications in areas such as plasma physics, electronics, biology, and optics \cite{drazin1989}. Mathematical description of solitons can be formulated in terms of model dependent non-linear partial differential equations (e.g., the non-linear Schr\"odinger equation). In general, it has been established that non-linear excitations (i.e., shock waves and solitons) exhibit distinct dependency between their amplitude and propagation velocity \cite{drazin1989}.

Solitons in thin $^4$He films adsorbed on solid substrates have been studied extensively by both experiments \cite{kono1981,hopkins1996,mckenna1990,levchenko1997} and theory \cite{huberman1978,nakajima1980,biswas1983,balakrishnan1990}. The film thickness is typically only a few atomic layers, which supports the propagation of third sound \cite{atkins1}. When the film is driven by a sufficiently large amplitude excitation, the response of the system becomes non-linear and typically follows the Korteweg-de Vries (KdV) equation \cite{mckenna1990,nakajima1980}. The KdV equation is known to support solitonic solutions, which has been confirmed experimentally for helium films in the previously mentioned references. Solitons have also been observed experimentally in related systems such as Bose-Einstein condensates (BEC) and $^3$He (magnetic solitons) \cite{burger1999,denschlag2000,anderson2001,khaykovich2002,strecker2002,nguyen2017,marchant2013,gould1976,mineyev1978,maki1977}. In the former case, experimental observations have been successfully modeled by the Gross-Pitaevskii (GP) equation  \cite{abdullaev2008,hamner2011,galitski2016}. However, bright solitons have not been observed in bulk superfluid $^4$He up to date. Such observation would not only provide important details of the underlying non-linear response of this quantum liquid, but it would also allow for the study of soliton dynamics (including dissipation) over much longer propagation distances and times than currently possible in BECs.

Studies of non-linear excitations in bulk superfluid helium are scarce. Most experiments have concentrated on the propagation of second sound shock waves \cite{osborne1951,iznankin1983,atkin1985} whereas non-linear first sound has received very little attention. In the latter case, the efforts have mainly concentrated on the construction of cryogenic compression shock tubes \cite{cummings1976,nellis1984,nagai2000}, which can be used to generate shock waves in the liquid and study their properties (e.g., velocity, amplitude). Shock waves, unlike solitons, are known to exhibit strong dissipation and dispersion \cite{remoissenet2003}. Semi-empirical analysis of shock waves can be carried out by the Rankine-Hugoniot theory or its extension that is applicable in the superfluid phase \cite{moody1984}. As shown in a recent study, shock waves in superfluid helium evolve on a nanosecond time scale and hence time-resolved experiments are required for their characterization \cite{garcia2016}.

Due to the lack of sufficiently accurate theoretical models for bulk superfluid helium, the possible existence of solitons and their properties in this medium have not been studied previously. Note that neither GP or KdV equations are applicable for superfluid helium. Only non-local phenomenological models, such as density functional theory (DFT) \cite{dalfovo,ancilotto2005}, can describe the atomic-scale static and dynamic response of superfluid helium accurately. While previous time-dependent DFT (TDDFT) calculations have noted the existence of supersonic non-linear waves \cite{eloranta2002,Her12}, their nature and properties were not studied further. Consequently, no experimental efforts have been put forward to prove (or disprove) the existence of solitons in this medium. The obvious differences between helium films and the bulk liquid are the dimensionality (2-D vs. 3-D) and the presence of the film supporting substrate. The latter influences the sound velocity as a function of depth, which makes the application of KdV-type equation attractive \cite{nakajima1980b}. 

In this work, we report on the first experimental observation of bright solitons in bulk superfluid $^4$He, which are created by rapidly expanding plasma and boiling on a metal target surface. In addition to the experimental evidence, their existence and dynamic properties are also studied by TDDFT.

\begin{figure}
\includegraphics[scale=0.5]{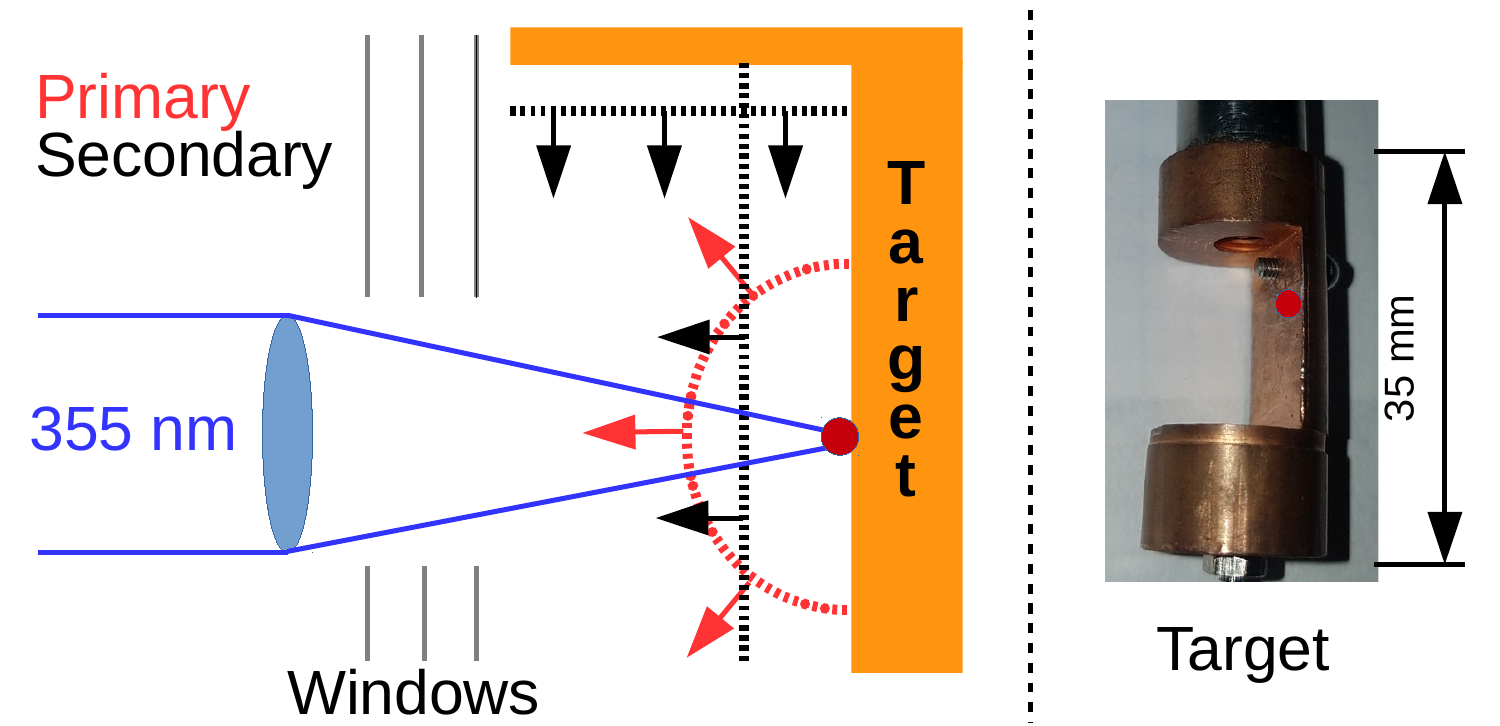}
\caption{Metal target geometry (scale accuracy $\pm 10$ $\mu$m). The primary waves (red) originate from the expanding plasma created by laser ablation (red circle) whereas the planar waves (black) are created by rapid boiling on the target surface. The optical axis for imaging is perpendicular to the plane shown. A photograph of the target is shown on the right.}
\label{fig1}
\end{figure}

The experiments employed a focused (spot diam. 50 $\mu$m) laser pulse (3rd harmonic 355 nm; 9 ns pulse length; 0.5 GW/cm$^2$; Continuum Minilite-II Nd-YAG laser) to generate plasma on the surface of a solid copper target immersed in bulk superfluid helium between 1.7 and 2.1 K at saturated vapor pressure (Oxford Variox or Janis 8DT cryostat) \cite{garcia2016}. A schematic target configuration is depicted in Fig. \ref{fig1}. The initial radial plasma expansion  \cite{buelna1,buelna2,sasaki1} leads to non-linear excitation of the surrounding liquid, which was visualized by time-resolved shadowgraph photography using a monochrome charge-coupled device (CCD; Imaging Source DMK23U445) equipped with 180X zoom lens (working dist. 95 mm; max. resol. 1.7 $\mu$m/pixel and focal depth $\pm$100 $\mu$m) and a delayed laser pulse (2nd harmonic 532 nm; 9 ns pulse length; Continuum Surelite-II Nd-YAG laser) as the background light. The contrast in the images is given by the Laplacian of the liquid density, which identifies the propagating wave edges. Due to scattering of the backlight, the images also show some contrast inside the wave.

\begin{figure}
\includegraphics[scale=0.55]{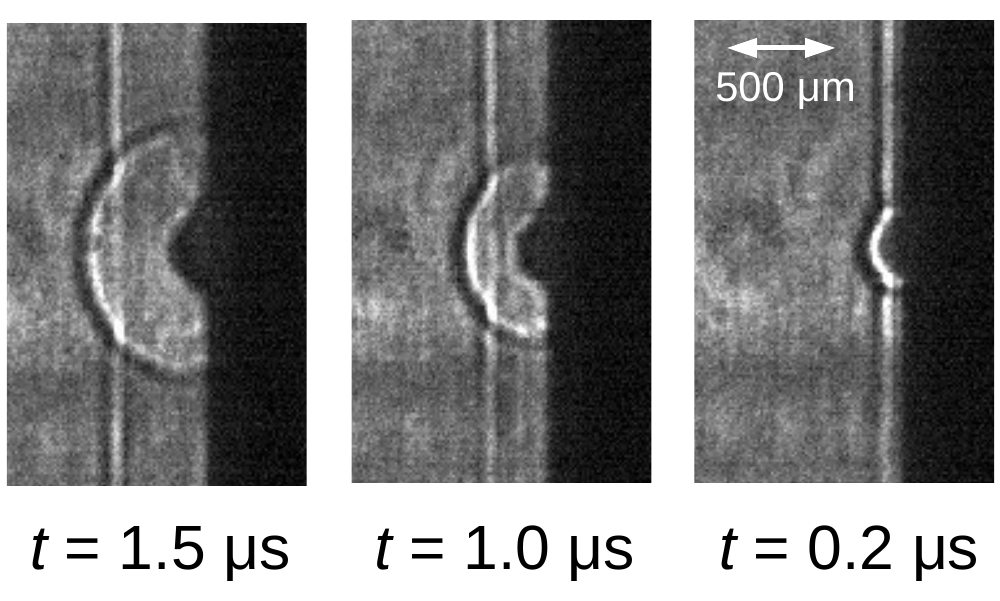}
\caption{Snapshots of the waves emitted from the target at given times $t$ ($T=1.7$ K). To increase the contrast in the images for display purposes, the lens system was placed slightly out of focus. The top portion of the target depicted in Fig. \ref{fig1} is not shown.}
\label{fig2}
\end{figure}

A closeup of the system at early times is shown in Fig. \ref{fig2}. The primary wave emission is produced directly by the expanding plasma (half-spherical geometry) whereas the secondary planar wave originates from boiling of liquid helium on the target surface and the subsequent rapid gas expansion. The latter process is a consequence of the fast heat transfer on the metal surface (propagation velocity up to 10$^5$ m/s) following the ablation event. In the long-time regime, both the primary half-spherical (width \textit{ca.} 15 $\mu$m) and secondary planar waves propagate in superfluid helium without dispersion (rate $<$ 0.025 m/s) until they disappear from the observation window after 10 mm. This behavior is consistent with solitons.

\begin{figure}
\includegraphics[scale=0.5]{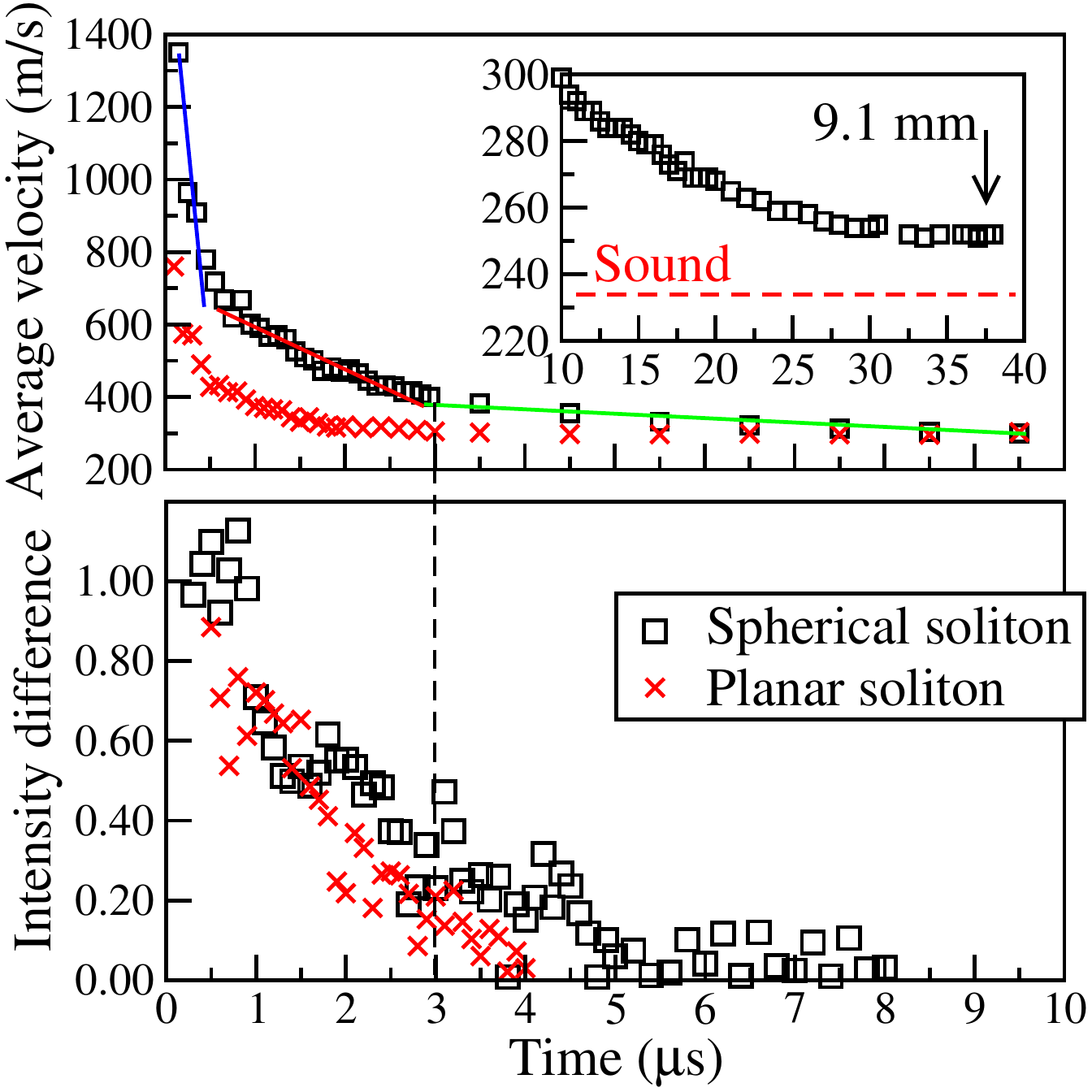}
\caption{The top panel shows the time evolution of the average soliton velocity (accuracy better than $\pm$ 1 m/s at long times) at 1.7 K with the long-time regime magnified in the inset (total travel distance indicated by an arrow). The three lines (blue, red, green) highlight the changes in the wave deceleration. The time axis refers to the delay between the ablation and backlight laser pulses. The bottom panel shows the normalized shadowgraph intensity difference in front of the wave vs. behind it. The dashed line provides a guide to the eye for correlating the velocity and intensity difference data.}
\label{fig3}
\end{figure}

The time evolution of normalized shadowgraph intensity difference in front of the soliton vs. immediately behind is shown in the bottom panel of Fig. \ref{fig3}. Assuming that the nature of the left-over liquid excitations (i.e., spatial variations in liquid density) does not evolve in time, this difference reflects the wave dissipation rate \cite{settles1}. During the first \textit{ca.} 3 $\mu$s, rapid dissipation of both spherical and planar solitons takes place along with the associated decrease in their propagation velocity (top panel of Fig. \ref{fig3}). The second regime ($t > 3$ $\mu$s) exhibits lower wave dissipation rate and the decrease in wave velocity begins to level off. As shown in the inset, the long-time ($t > 30$ $\mu$s) soliton propagation appears nearly dissipationless as the velocity remains constant at slightly above the speed of first sound (instantaneous velocity \textit{ca.} 250 m/s; see Ref. \cite{garcia2016}). This limiting velocity follows the same temperature dependence as the first sound. We attribute the fast initial dissipation and reduction in the propagation velocity to wave crest breaking process where the high density liquid is left behind as shocks. Note that a small decay in the velocity is also expected due to the finite viscosity present in the experiments \cite{huberman1978} and the change in volume of the spherical soliton with increasing radius.

\begin{figure}
\includegraphics[scale=0.55]{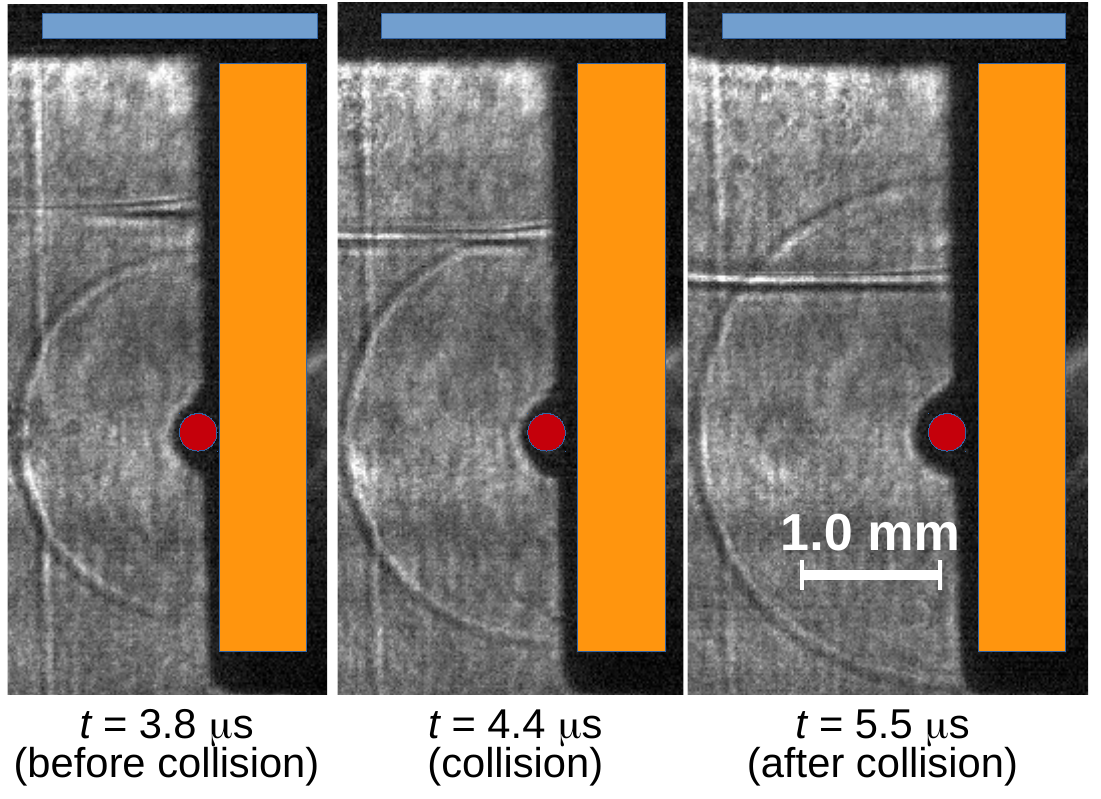}
\caption{Time evolution of the primary (spherical) and secondary (planar from top) solitons before, during, and after collision ($T = 1.7$ K). The red circle indicates the origin for the primary wave emission and the top (blue) surface for the secondary wave. The delay between the ablation and backlight laser pulses is indicated by $t$.}
\label{fig4}
\end{figure}

Another inherent property of solitons is that they emerge from a two-wave collision without any apparent change to their shape (apart from a possible change in phase). Collision between two solitons is shown in Fig. \ref{fig4}, where the primary half-spherical wave collides with the planar wave originated from the top section of the target. Due to the geometry of the emitted waves (half sphere vs. plane), the two waves must intersect at the focal plane of the imagining system. The shadowgraph images clearly show that the solitons do not interact and continue to propagate unchanged after the collision. Furthermore, collision of the solitons with a metal surface (not shown) leads to effective reflection, but this is accompanied by energy loss as evidenced by an audible mechanical shock emitted into the metal. In the long-time regime, the reflected solitons from the cryostat walls can be observed to reach the target region again (total travel dist. 10 cm).

In addition to the experimental observations discussed above, we have also carried out TDDFT calculations in 3-D \cite{dftreview}  to identify solitonic solutions in bulk superfluid $^4$He and study their dynamic properties. Within this model, helium is described by a complex valued order parameter $\Psi( \mathbf{r},t)$, which is related to the atomic density as $\rho (\mathbf{r},t)= |\Psi( \mathbf{r},t)|^2$. The TDDFT equation is 
\begin{equation}
\imath \hbar\frac{\partial}{\partial t}  \Psi({\mathbf r},t)  = \left\{
-\frac{\hbar^2}{2m}\nabla^2 + \frac{\delta{\cal E}_{c}}{\delta\rho}
\right\} \,\Psi(\textbf{r},t) 
\label{eq1}
\end{equation}
where $m$ is the mass of $^4$He and the functional ${\cal E}_c\left[\rho\right]$ was taken from Ref.  \onlinecite{ancilotto2005}. This functional includes both finite-range and non-local corrections that are required to describe the $T=0$ response of liquid $^4$He accurately on the \AA{}ngstr\"om-scale. Note that this model does not include viscous dissipation and cannot be propagated over long times (microseconds) due to limitations in current computational resources. For this reason, TDDFT cannot be used to study the related dissipative effects observed in the experiments. Furthermore, the accessible length scale is also very different from the experiments (i.e., nm vs. $\mu$m). However, as discussed below, the TDDFT results can be scaled up to match the experiments.

\begin{figure}
\includegraphics[scale=0.5]{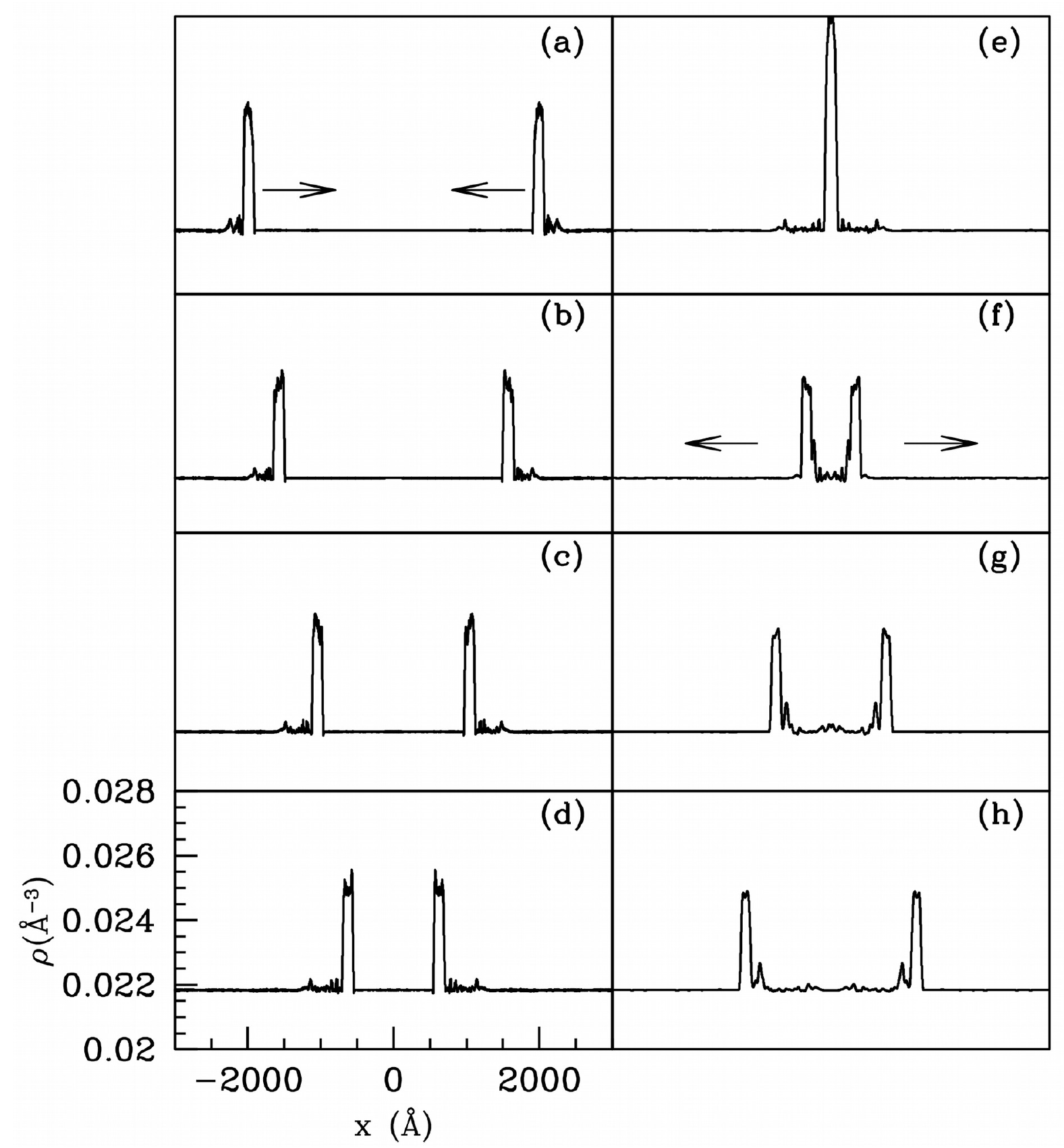}
\caption{Snapshots of superfluid $^4$He density along the direction of soliton propagation ($x$-axis) from TDDFT. The initial compression was created on the simulation box boundary with $n = 30$ (see text). Due to periodic boundary condition, the solitons propagate as shown in (a) - (d), collide in (e), and continue propagating almost unchanged in (f) - (h). When compared with experiments, the $x$-axis is oriented perpendicular to the ablation target.}
\label{fig5}
\end{figure}

To mimic the initial condition in the experiments (i.e., sudden compression by expanding plasma), the initial order parameter, $\Psi({\bf r},0)=\sqrt{\rho ({\bf r},0)}$, is constructed as
\begin{equation}
\rho ({\bf r},0)= \rho_0 [1 + (\Delta \rho /\rho_0) \sin^2(\pi x/\lambda _c)\Theta_w(x)]
\label{eq2}
\end{equation}
where $\Theta_w$ represents a ``box function" centered at $x_0$ with width $w=n\lambda_c$ (with $n$ integer), i.e., $\Theta_w(x)=1$ when $x_0-w/2<x<x_0+w/2$ and $\Theta _w=0$ otherwise. Eq. (\ref{eq2}) represents a square profile with average value $\Delta \rho/2$ that is superimposed on uniform bulk density $\rho _0$ ($0.0218\,$\AA$^{-3}$ at $T = 0$) and modulated along the $x$-axis with wavelength $\lambda _c=3.58\,$\AA{}. This ansatz is based on the following assumptions: 
(i) when the liquid is rapidly compressed, the local density is increased with respect to the bulk and (ii) liquid $^4$He flowing at a velocity greater than the Landau critical velocity ($v_L$) undergoes a transition from a spatially homogeneous liquid to a layered state characterized by a periodic density modulation along the direction of propagation (wavelength $\lambda_c$ and amplitude determined by $v-v_L$) \cite{Pit84,Anc05}. Such layered structures with densities higher than the bulk have also been observed in DFT simulations of fast moving particles in liquid $^4$He \cite{ancilotto2017}.

During the early stages of the time evolution of $\rho({\bf r},t)$, dispersive low-amplitude supersonic waves with wavelength $\sim \lambda _c$ were produced (not shown). For the sake of clarity, we show smoothed density profiles after taking a local average of the density within a space window of $\pm 2\lambda_c$. We wish to stress that this procedure is just post-processing and therefore it does not affect the time evolution itself. Note that the applied theoretical model must be able to describe the underlying atomic scale internal structure of the soliton. 

When the initial state given by Eq. (\ref{eq2}) is propagated in time using Eq. (\ref{eq1}), it splits rapidly into two counter propagating bright solitons as shown in Fig. \ref{fig5}. In contrast, due to the presence of the expanding plasma and the target in the experiments, only one soliton may form following the initial compression. The initial position $x_0$ of the square profile was placed at the simulation box boundary and, due to the periodic boundary condition, the two solitons resulting from the initial splitting move away from the boundaries towards the center. Note that the soliton width ($\sim 20$ nm for the case shown) and height are well preserved during the time evolution. In comparison, a gaussian wave packet with the same width and amplitude would disperse rapidly at 90 m/s. The solitons were also found to be stable with respect to random distortions introduced into the order parameter.

\begin{figure}
\includegraphics[scale=.43]{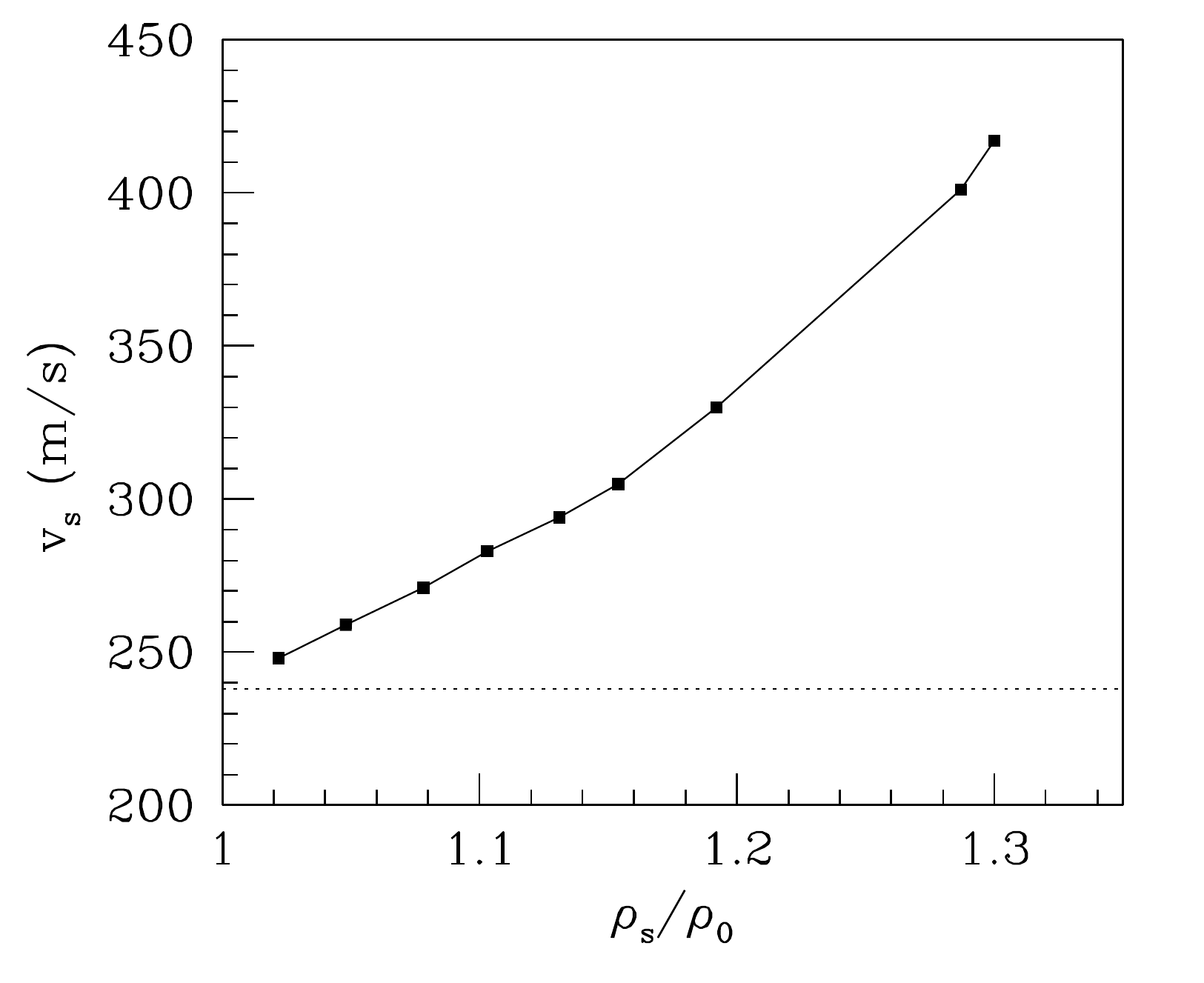}
\caption{Soliton propagation velocity ($v_s$) vs. amplitude ($\rho_s/\rho_0$) from TDDFT. The horizontal dotted line indicates the velocity of first sound in superfluid $^4$He at $T = 0$.}
\label{fig6}
\end{figure}

The relationship between the average soliton height $\rho_s$, which is controlled by the value of $\Delta \rho/\rho_0$ in Eq. (\ref{eq2}), and its propagation velocity exhibits nearly linear behavior at low amplitudes as shown in Fig. \ref{fig6}. In the limit of very small amplitudes, the velocity approaches the speed of sound. When the amplitude $\rho_s/\rho_0$ is increased above 1.3, the system becomes unstable and evolves rapidly into a series of shock waves. This instability may be related to the previously mentioned wave crest breaking phenomenon, which was observed before 3 $\mu$s in Fig. \ref{fig3}. The calculated maximum stable soliton velocity (\textit{ca.} 430 m/s) corresponds approximately to the point where the rapid velocity decay levels off (dashed line near 3 $\mu$s in Fig. \ref{fig3}). The long-time wave propagation velocity in the experiments remained slightly above the speed of sound, which corresponds to \textit{ca.} 3\% density increase at the soliton with respect to the bulk liquid (cf. Figs. \ref{fig3} and \ref{fig6}). Both the lack of dispersion and the distinct amplitude-velocity dependence are characteristic to solitons.

When the initial width of the compression in Eq. (\ref{eq2}) is increased, the width of the emitted solitons increases accordingly. Therefore, despite of the obvious difference in the length scale between TDDFT and the experiments, this suggests that the presented nanometer-scale mechanism can scale up to micrometers. We also note that the presented solitonic waves from Eq. (\ref{eq2}) can only be observed using a finite-range non-local energy density functional whereas local models, such as GP fitted to reproduce the speed of sound, do not support such solutions.

A collision between two solitons from TDDFT is shown in Fig. \ref{fig5}. Based on the simulations, the amplitude, shape, and velocity of the solitons are well preserved after the collision. This observation is in agreement with the experimental images shown in Fig. \ref{fig4}. At the point of collision shown in panel (e), the solitons interfere constructively as they both have a common phase factor (i.e., identical origin). If a soliton is made to collide with an exponentially repulsive wall (not shown), TDDFT calculations show that it loses its shape partially and dissipates some of the energy as shock waves. This behavior is also consistent with the experimental observations.

In summary, we have shown for the first time that bulk superfluid $^4$He can support bright solitonic waves. This is evidenced by both direct experimental observations as well as theoretical modeling based on TDDFT. The liquid compression created by the expanding plasma is sufficiently high such that the resulting non-linear response can counteract the dispersive effects. This is in contrast to previously studied thin liquid helium films where the presence of the supporting substrate played a major role in producing the necessary non-linear response. In bulk superfluid helium, solitons become unstable when their amplitude exceeds a critical threshold, which corresponds to a velocity slightly above 400 m/s.

\begin{acknowledgments}
This work was supported by National Science Foundation grant DMR-1205734. The authors thank M. Barranco, M. Pi, and L. Salasnich for helpful discussions.
\end{acknowledgments}

\end{document}